\documentclass[twocolumn,showpacs,preprintnumbers,amsmath,amssymb]{revtex4}

\usepackage{graphicx}
\usepackage{dcolumn}
\usepackage{bm}
\usepackage{epsfig}

\begin{document}
%
%
\title{Nonclassical imaging for a quantum search of trapped ions}
\author{G.S. Agarwal$^{1,2}$}
\author{G.O. Ariunbold$^{1,3}$}%
\author{J. von Zanthier$^{1,4}$}
\author{H. Walther$^{1,4}$}
\affiliation{$^{1}$Max-Planck-Institut f${\rm \ddot{u}}$r
Quantenoptik, M${\rm \ddot{u}}$nchen, D-85748 Garching, Germany
\\
$^{2}$Physical Research Laboratory, Navrangpura, Ahmedabad-380
009, India
\\
$^{3}$Theoretical Physics Laboratory, National University of
Mongolia, 210646 Ulaanbaatar, Mongolia
\\
$^{4}$Sektion Physik der LMU, M${\rm \ddot{u}}$nchen, D-85748
Garching, Germany}
\date{June 4, 2004}
\begin{abstract}
We discuss a simple search problem which can be pursued with
different methods, either on a classical or on a quantum basis.
The system is represented by a chain of trapped ions. The ion to
be searched is a member of that chain, consists, however, of an
isotopic species different to the others. It is shown that the
classical imaging may lead as fast to the final result as the
quantum imaging. However, for the discussed case the quantum
method gives more flexibility and higher precision when the number
of ions considered in the chain is increasing. In addition,
interferences are observable even when the distances between the
ions is smaller than half a wavelength of the incident light.
\end{abstract}
\maketitle
Quantum search algorithms \cite{grover} enable us to determine an
object from a black box with $N$ elements by a number of
measurements which is of the order of $\sqrt{N}$. A quantum search
thus provides a polynomial speed-up in comparison with any known
classical algorithm. Several methods have been proposed for
implementing a quantum search \cite{Chuang98}. In this letter we
discuss a very simple example where the search can be performed in
the same system either on a classical or a quantum basis. Such a
situation is useful for illustrating the particular aspects of a
quantum search with respect to a classical search. We recall that
fluorescence imaging, i.e., measurement of the mean intensity of
radiation scattered by atoms upon excitation, provides information
about the density profile. In fact, this is a common method used
to image, for example, trapped ions or a Bose-Einstein condensate
\cite{Walther92,Blatt98,Ketterle00}. We could, however, do more
than just image the sources of fluorescence, e.g. by determining
the spectrum of the scattered light. In this case, in addition to
the position the motional state of the atoms is accessible.
Instead of imaging the particles individually on a detector, one
could also observe the fluorescence in the far field or Fourier
plane of a lens \cite{eichmann}. In this case, the intensity
distribution corresponds to the interference pattern described by
the first order correlation function of the fluorescence light. It
results from the contribution of all particles at once, and their
relative position can be deduced from the interference pattern.
The intensity distribution thus contains more information than the
direct imaging of the ions \cite{grover2}. A further step in
sophistication would be to observe the scattered light using two
detectors and to measure the second order correlation function.
This corresponds to a nonclassical imaging technique where the
associated spatial distribution is determined by the different
paths the photons can take when reaching the two detectors; the
corresponding interference pattern again relies on the
contribution of all scatterers simultaneously but for certain
excitation angles is purely due to quantum interferences
\cite{skornia01,agarwal02}. In this type of experiment only the
coincidence events at the two detectors are recorded. As the paths
of the photons contributing to these coincidences change with the
relative position of the detectors and/or the scatterers, the
observed interference patterns again allow the positions of the
individual particles to be retraced. However, owing to the larger
variability of the parameters involved, a much richer interference
structure is obtained. To illustrate this in more detail we
consider a particular example, viz. a linear chain of ions of the
same atomic species. In order to set up a search procedure we
assume that one of the ions belongs to an isotopic species
different to the others. By calculating the spatial photon -
photon correlations we then show how the second order correlations
can be used to reveal information on the position of the
off-resonant, non-radiating isotope in the chain. We will also
demonstrate why this information is more extensive than that
obtained from the usual fluorescence imaging techniques or first
order correlation function. In particular, it will be shown that
the second order correlation function is able to provide data in
parameter regions where the first order correlation function is
not able to provide any information.
\par
Let us consider a chain of ions with an energy
level scheme as shown in Fig.~1a. The transition $|
g\rangle\rightarrow| e\rangle$ is used for exciting the system,
whereas the transition $| e\rangle\rightarrow | f\rangle$ serves
for fluorescence detection. The excitation could be performed, by
for example, a short pulse of radiation. The most basic search
protocol would consist in resonantly exciting each ion one by one
on the $| g\rangle \rightarrow | e\rangle$ transition using a
focused laser beam and observing the fluorescence scattered by the
ion on the $| e\rangle\rightarrow| f\rangle$ transition. With the
isotope being off-resonant and remaining dark upon excitation, it
will take at most $N-1$ steps to localize the isotope, i.e. in the
mean $N/2$ trials. One could next observe the fluorescence
intensity in the far field or in the Fourier plane of a lens, i.e.
without imaging the ions. The corresponding interference pattern -
produced by the simultaneous superposition of the electromagnetic
fields of all scatterers - can be used to extract information
about the position of the isotope. To be explicit, let us
calculate the far-field intensity produced by a chain of ions at
some distance $r$. The positive frequency part of the field
amplitude can be written as (see \cite{agarwalbook})
\begin{eqnarray} \label{efield}
\vec E^{(+)} (\vec r_1,t)  & = & \vec E^{(+)}_0 (\vec r_1,t)
\\
& - & \frac{k^2 e^{i  k r}}{r} \sum_{j } {\rm e}^{-i k \vec n_1
\vec R_j} \vec n_1 \times (\vec n_1 \times \vec d _{f e})
A^{(j)}_{f e},\nonumber
\end{eqnarray}
where $\vec n_1$ stands for a unit vector in the direction of the
detector position, $\vec r_1 = \vec n_1 r$, $k = \omega_0 /c$.
$\omega_0$, $\vec d_{f e}$ and $A^{(j)}_{f e} = | f\rangle_{j j}
\langle e |$ are, respectively, the transition frequency, dipole
matrix element and the dipole operator of ion $j$ for the
transition $| e\rangle\rightarrow {| f\rangle}$ and the sum is
over all ion positions $\vec R_j$ in the chain. The far-field
intensity is thus determined by the first order correlation
function $G^{(1)}(\vec r_1)= \langle A^{\dagger} (\vec r_1) A(\vec
r_1) \rangle$, where $A(\vec r_1) = \sum_{j } A^{(j)}_{f e} {\rm
e}^{-i k \vec n_1\vec R_j}$. In the case of uncorrelated ions ($
\langle A^{(j)}_{e f}A^{(i)}_{f e}\rangle =\langle A^{(j)}_{e
f}\rangle\langle A^{(i)}_{f e}\rangle$), $G^{(1)} (\vec r_1 )$ can
be simplified to \cite{agarwalbook}
\begin{equation} \label{intensity1}
G^{(1)}(\vec r_1 ) = \sum_{j } \langle A^{(j)}_{e e} \rangle +
\sum_{j \not= i} \langle A^{(j)}_{e f} \rangle \langle A^{(i)}_{f
e} \rangle {\rm e}^{i k \vec n_1(\vec R_i - \vec R_j)}.
\end{equation}
According to (\ref{intensity1}), the information on the spatial
structure of the ion chain is contained in $G^{(1)} (\vec r_1 )$
as long as $\langle A^{(j)}_{e f} \rangle \not= 0$. This is the
case as long as, for example, no $\pi$-pulse is used for the
excitation. The information about the localization of a
non-radiating particle is also contained in (\ref{intensity1}), as
$G^{(1)} (\vec r_1 )$ will look different for different positions
of the isotope. This will be discussed in more detail below. Note,
however, that a result similar to (\ref{intensity1}) is obtained
also in the case of classical dipoles or classical antennas, i.e.
in the case of classical light sources \cite{grover2}.
\\ \par
As mentioned in the introduction, we could, however, do more than
just measure the far-field intensity, e.g. by examining the
quantum features of the emitted radiation. For that purpose let us
study the spatial photon - photon correlations produced by the
chain of ions by using two photodetectors (see Fig. 1b).
\begin{figure}[!ht]
\begin{center}
\includegraphics[width=60mm]{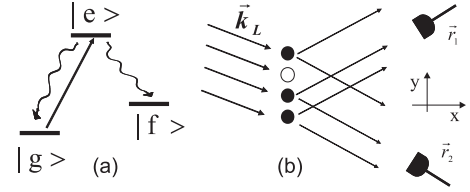} \label{arifig1}
\caption{(a) Level scheme considered for the trapped ions: the
ions are excited on the $|g\rangle \rightarrow |e \rangle$
transition, whereas the $|e\rangle \rightarrow |f \rangle$
transition is used for fluorescence detection; (b) an incident
$\pi$-pulse of a wave with wave vector $\vec k_L$ excites the
linearly trapped ions into their upper level $|e\rangle$. The
fluorescence is then registered in the far field by two detectors
at $\vec r_{i}$, $i=1,2$ ($|\vec r_1|=|\vec r_2|=r$).}
\end{center}
\end{figure}
The photon - photon correlations are determined by the expression
(see \cite{agarwalbook})
\begin{equation} \label{G2}
\langle\vec E^{(-)} ( \vec r_1 , t )\vec E^{(-)} ( \vec r_2 , t
)\vec E^{(+)} ( \vec r_2 , t )\vec E^{(+)} ( \vec r_1 , t
)\rangle,
\end{equation}
where $\vec r_2=\vec n_2 r$ defines the position of the second
detector and $\vec E^{(-)} ( \vec r_i, t )$ denotes the complex
conjugate of $\vec E^{(+)} ( \vec r_{i} , t )$, $i=1,2$. As can be
seen from (\ref{G2}), the spatial photon - photon correlations are
determined in terms of the atomic/ionic operators through
\begin{equation} \label{G2a}
G^{(2)} ( \vec r_1 , \vec r_2 ) = \langle A^{\dagger} (\vec r_1)
A^{\dagger} (\vec r_2) A(\vec r_2) A (\vec r_1) \rangle.
\end{equation}
In order to demonstrate how the information on the spatial
structure of the chain is contained in the quantity $G^{(2)}( \vec
r_1, \vec r_2 )$, let us first examine the case of two ions. If
the ions are initially prepared in the state $| e\rangle$, we find
(see \cite{agarwal02})
\begin{equation} \label{G2twoions}
G^{(2)} ( \vec r_1 , \vec r_2 ) = B ( 1 + \cos ( k (\vec n_2 -
\vec n_1 )( \vec R_A - \vec R_B ) ) ),
\end{equation}
where $B$ is a constant. Obviously, the information on the
location of the two ions is contained in the nonclassical
\cite{agarwal02} interference pattern $G^{(2)} ( \vec r_1 , \vec
r_2 )$ via the atomic position variables $\vec R_A$ and $\vec
R_B$. Let us consider next a chain of $N$ ions of which one is an
off-resonant, non-radiating isotope. The explicit calculation in
case of $\pi$-excitation leads to the following result:
\begin{equation} \label{G2general}
G^{(2)}_{p,N} ( \vec r_1 , \vec r_2 ) =\sum_{i,j\neq p (i<j)}
   |\gamma_{ij}( \vec r_1 , \vec r_2 )|^2,
\end{equation}
where $p=\overline{1,N}$ stands for the position of the isotope,
$i=\overline{1,N-1}$ and $j=\overline{2,N}$ and $\gamma_{lm} (
\vec r_1 , \vec r_2 ) =\alpha_l( \vec r_1)\beta_m( \vec r_2
)+\alpha_m( \vec r_1 )\beta_l( \vec r_2 )$ with $\alpha_{l} ( \vec
r_1 ) ={\rm e}^{i k {\vec n}_{1}{\vec R}_{l}}$ and $\beta_{m} (
\vec r_2 ) ={\rm e}^{i k {\vec n}_{2}{\vec R}_{m}}$ ($l, m =
\overline{1,N}$). The solution (\ref{G2general}) can easily be
interpreted as the sum of terms associated with all possible
optical path differences between the photons when scattered by two
different ions and recorded by the two detectors, on the
assumption that the isotope at $p$ does not scatter at all.
Equation (\ref{G2general}) becomes even more transparent for
equally spaced ions. In the case of four ions, for example, the
system after initial excitation by a $\pi$-pulse will be in one of
the pure states $| f e e e\rangle$, $| e f e e\rangle$, $| e e f
e\rangle$ or $| e e e f\rangle$. In this case we get
$|\gamma_{12}( \vec r_1 , \vec r_2 )|^2 \equiv |\gamma_{23}( \vec
r_1 , \vec r_2 )|^2 \equiv |\gamma_{34}( \vec r_1 , \vec r_2 )|^2
= 2+2 c_1( \vec r_1 , \vec r_2 )$ and $|\gamma_{13}( \vec r_1 ,
\vec r_2 )|^2 \equiv |\gamma_{24}( \vec r_1 , \vec r_2 )|^2 =
2+2c_2( \vec r_1 , \vec r_2 )$, where $c_{m}( \vec r_1 , \vec r_2
)=\cos[k(\vec n_1-\vec n_2)(\vec R_{1} -\vec R_{m+1})]$, so that
\begin{eqnarray} \label{aricorr4}
&& G^{(2)}_{p,4} ( \vec r_1 , \vec r_2 )=
\\ \nonumber
&& = 2(\delta_{p,1} + \delta_{p,4}) (3 + 2 c_1 ( \vec r_1 , \vec
r_2 ) +c_2 ( \vec r_1 , \vec r_2 ) )
\\ \nonumber
&& + 2(\delta_{p,2} + \delta_{p,3})(3+c_1( \vec r_1 , \vec r_2
)+c_2( \vec r_1 , \vec r_2 )+c_3( \vec r_1 , \vec r_2 )).
\end{eqnarray}
where $\delta_{p,q}$ is the Kronecker symbol.
\begin{figure}[!ht]
\begin{center}
\includegraphics[width=90mm]{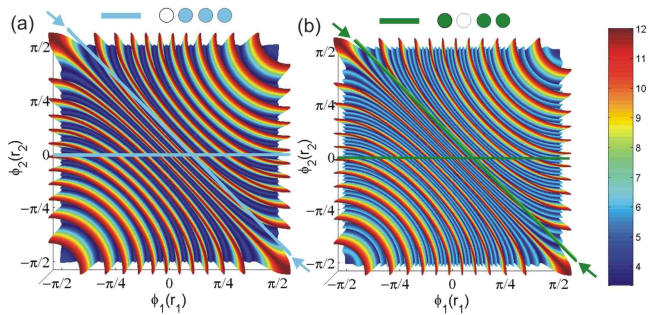} \label{arifig2}
\caption{3-D plot of the second order correlation function
$G^{(2)}_{p,4} ( \vec r_1 , \vec r_2 )$ versus the two detector
positions $\phi_{i}(\vec r_i),$ $i=1,2$, for $N=4$ ions, with the
off-resonant, non-radiating isotope placed in (a) the first (or
fourth) position and (b) the second (or third) position of the
chain. The distance between neighboring ions is $d=5.75 \lambda$,
wavelength of the incident light is $\lambda=194$nm and a $\pi$
pulse is used for the excitation.}
\end{center}
\end{figure}
\begin{figure}[!ht]
\begin{center}
\includegraphics[width=78mm]{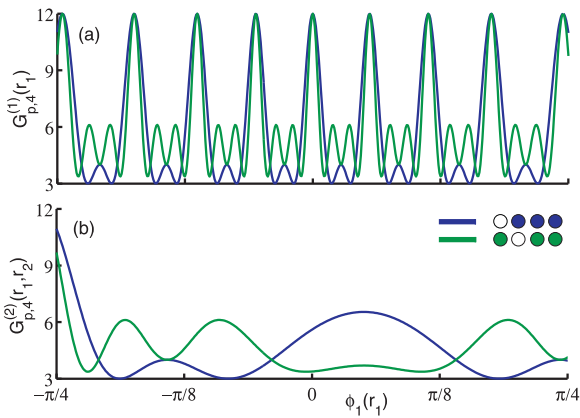} \label{arifig3}
\caption{Interference patterns for $N=4$ ions. Blue and green
curves correspond to the first (or fourth) and second (or third)
ion being the isotope, respectively. Parameters are the same as in
Fig. 2. (a) Classical interference pattern $G^{(1)}_{p,4} ( \vec
r_1)$ (or second order correlation function $G^{(2)}_{p,4} ( \vec
r_1 , \vec r_2 )$ where the second detector is at $\phi_2(\vec
r_2)=0$). (b) Nonclassical interference pattern $G^{(2)}_{p,4} (
\vec r_1 , \vec r_2 )$; the two detectors are slightly separated
with $||\phi_1|-|\phi_2||=1/\pi$.}
\end{center}
\end{figure}
\begin{figure}[!ht]
\begin{center}
\includegraphics[width=80mm]{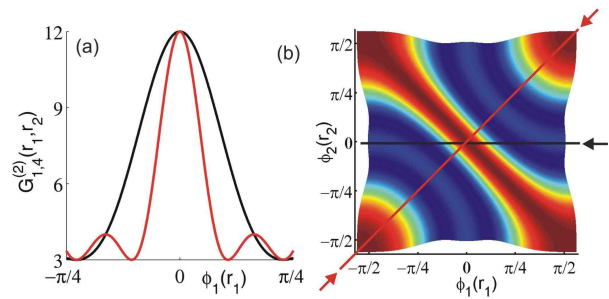} \label{arifig4}
\caption{Interference patterns for $N=4$ ions with ion distance
$\lambda/2$. (a) Red curve: nonclassical interference pattern
$G^{(2)}_{1,4} ( \vec r_1 , \vec r_2 )$ with $\phi_1(\vec
r_1)=\phi_2(\vec r_2)$ (see red line in (b)), black curve:
classical interference pattern $G^{(1)}_{1,4} ( \vec r_1)$
($G^{(2)}_{1,4} ( \vec r_1 , \vec r_2 )$ with $\phi_2(\vec
r_2)=0$; see black line in (b)).}
\end{center}
\end{figure}
\begin{figure}[!ht]
\begin{center}
\includegraphics[width=78mm]{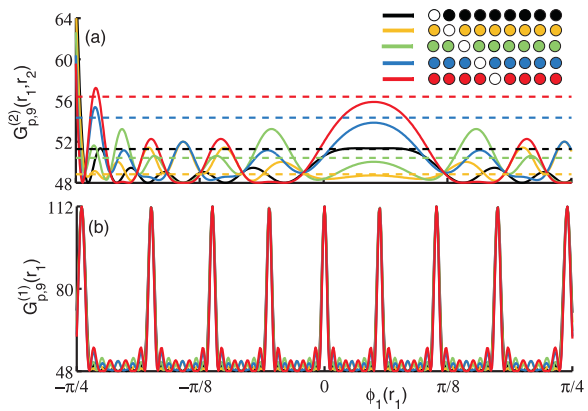} \label{arifig5}
\caption{Interference patterns for $N=9$ ions. Black, yellow,
green, blue and red curves correspond to 1st (or 9th), 2nd (or
8th), 3rd (or 7th), 4th (or 6th) and 5th ion being the isotope,
respectively. Parameters are the same as in Fig. 2. (a)
Nonclassical interference pattern $G^{(2)}_{p,9} ( \vec r_1 , \vec
r_2 )$; the two detectors have a constant separation
$||\phi_1|-|\phi_2||=1/\pi$. Dashed lines correspond to detectors'
positions being chosen in such a way that $|sin|\phi_1| -
sin|\phi_2||=0.378$. (b) Classical interference pattern
$G^{(1)}_{p,9} ( \vec r_1)$.}
\end{center}
\end{figure}
According to (\ref{aricorr4}), the information about the isotope
position can be extracted from $G^{(2)}_{p,4} ( \vec r_1 , \vec
r_2 )$ due to the unique distribution of prefactors (or Fourier
coefficients) $c_j$ for the different isotope localizations. From
(\ref{aricorr4}) it is seen, however, that the outcome for the
isotope positions $p = 1$ ($2$) is identical to the outcome for $p
= 4$ ($3$). This means that for equally spaced ions an additional
measurement would be required to determine whether the isotope is
on the left or right hand side with respect to the middle of the
chain. A similar argument holds for the general outcome
(\ref{G2general}): for arbitrarily spaced ions there is a unique
interference pattern for each isotope position $p$ since the
$\gamma_{ij}$ will in general be different for different $i, j$,
so that there is a unique combination of $\gamma_{ij}$ for each
isotope position $p$. By measuring the spatial dependence of the
photon-photon correlation function one could thus clearly
distinguish the isotope position $p$ from any other position $p'$.
Note in particular that the additional degree of freedom given by
$\vec r_2$ due to the use of two detectors increases the parameter
space available. This is exceedingly useful when the information
from the first order correlation function $G^{(1)}_{p,N}( \vec r_1
)$ is difficult to extract. This is the case for, for example, a
large number of ions and/or a small ion spacing. There is in
addition the particular situation where $G^{(1)}_{p,N}( \vec r_1
)$ contains no information at all, as is the case, for example,
for a $\pi$-pulse excitation. In this case the contrast of the
interference pattern of $G^{(1)}_{p,N}( \vec r_1 )$ vanishes,
whereas that of $G^{(2)}_{p,N} ( \vec r_1 , \vec r_2 )$ can still
remain maximal \cite{skornia01,agarwal02}. In Fig. 2, we show the
behavior in case of four equally spaced ions for $G^{(2)}_{p,4} (
\vec r_1 , \vec r_2 )$ for $p = 1$ (or $4$) and $p=2$ (or $3$) as
a function of $\phi_{i} (\vec r_{i}) = \arctan( r_{i,y} /
r_{i,x})$, $i=1,2$. For the same excitation angle as used in this
figure (i.e. $\pi$-pulse excitation) $G^{(1)}_{p,4} ( \vec r_1 )$
would correspond to a constant, independently of $\phi_1(\vec
r_1)$ and independent of the isotope position $p$. For a
$\pi/2$-pulse excitation on the $|f\rangle \rightarrow |e\rangle$
transition, however, $G^{(1)}_{p,4} ( \vec r_1 )$ would - apart
from a prefactor - show an interference pattern similar to
$G^{(2)}_{p,4} ( \vec r_1 , \vec r_2 )$ for $\pi$-pulse excitation
and $\phi_2 (\vec r_2) = 0$ (see horizontal lines in Fig. 2). More
precisely, we obtain
\begin{eqnarray} \label{ari1corr5}
G^{(1)}_{p,4} ( \vec r_1 )&=& 1/2 (\delta_{p,1} + \delta_{p,4}) (3
+ 2 c_1 ( \vec r_1 ) + c_2 ( \vec r_1 ) )
\\ \nonumber
& + & 1/2 (\delta_{p,2} + \delta_{p,3})(3 + c_1( \vec r_1 ) + c_2(
\vec r_1 ) + c_3( \vec r_1 )).
\end{eqnarray}
where $c_{m}( \vec r_1 )=\cos[k \vec n_1(\vec R_{1} -\vec
R_{m+1})]$ and $p=\overline{1,4}$. The result (up to a prefactor)
is shown in Fig. 3a. Yet, it is here that the additional degree of
freedom of $G^{(2)}_{p,4} ( \vec r_1 , \vec r_2 )$ with respect to
$G^{(1)}_{p,4} ( \vec r_1 )$ becomes important: the additional
parameter $\vec r_2$ increases the available parameter space and
in this manner allows a more flexible and precise search of the
isotope. To show that, we plot in Fig. 3b $G^{(2)}_{p,4} ( \vec
r_1 , \vec r_2 )$ for $p = 1$ (or $4$) and $p= 2$ (or $3$), and
with $\vec r_1$ and $\vec r_2$ such that $| | \phi_1 (\vec r_1) |
- | \phi_2 (\vec r_2) | | = 1/\pi$. The latter condition stands
for a fixed distance between the two detectors and corresponds in
Fig. 2 to a straight line in the $(\phi_1 (\vec r_1), \phi_2(\vec
r_2))$-plane (see tilted lines in Fig. 2). As is seen from Fig.
3b, the constraints for the angular resolving power are much more
relaxed in comparison with that needed for $G^{(1)}_{p,4}(\vec
r_1)$ (Fig. 3a).
\par
The spatial second order correlation function also allows the
isotope position to be determined in cases where the ions are
separated by only $\lambda/2$ so that they cannot be individually
resolved. This can be demonstrated by analyzing $G^{(2)}_{1,4} (
\vec r_1 , \vec r_2 )$ for equally spaced ions with ion separation
$d = \lambda/2$ (Fig. 4). If the detector positions are chosen
such that $\phi_1 (\vec r_1) = \phi_2 (\vec r_2)$ - corresponding
in the $(\phi_1(\vec r_1),\phi_2(\vec r_2))$-plane to a straight
line with slope 1 - we resolve for $G^{(2)}_{1,4} ( \vec r_1 ,
\vec r_2 )$ a central maximum plus two side maxima in the central
region $-\pi/4 \leq \phi_1 (\vec r_1) \leq \pi/4$, whereas for
$G^{(1)}_{1,4} ( \vec r_1 )$ only the central maximum is obtained
(Fig. 4a). However, it is from the position and amplitude of the
side maxima that the information about the isotope position $p$ is
derived, whereas the central maximum in this regard contains no
information.
\par
Let us finally consider nine equally spaced ions in the trap one
of which is an isotope. There are $5$ different positions for the
isotope which can be distinguished from $G^{(2)}_{p,9}( \vec r_1 ,
\vec r_2 )$ (see above). These five different possibilities are
plotted in Fig. 5a on the assumption that $| | \phi_1 (\vec r_1) |
- | \phi_2 (\vec r_2) | | = 1/\pi$ and a $\pi$-pulse is used for
excitation. It can be clearly seen from the figure that the
requirement for the angular resolving power is far less demanding
in comparison with Fig. 5b, where $G^{(1)}_{p,9}( \vec r_1)$
(apart from a prefactor) is plotted for a $\pi/2$-pulse excitation
on the $|f\rangle \rightarrow |e\rangle$ transition. For a
relative position of the two detectors such that $|| \sin \phi_1
(\vec r_1) | - |\sin \phi_2 (\vec r_2) ||=0.378$, one even obtains
for $G^{(2)}_{p,9} ( \vec r_1 , \vec r_2 )$ constant values, with
amplitudes which depend on the isotope position $p$, but are
independent from $\phi_1(\vec r_1)$. These values are depicted in
Fig. 5a by dashed lines. However, due to the nonlinear dependence
of the two detector positions, this situation might be more
difficult to implement experimentally. Nevertheless, such as
special angular independent case could play an important role when
one would be interested to normalize the correlation function,
e.g. for the purpose of a better comparison between the different
isotope positions.
\par
In conclusion, it has been shown in a simple search problem using
a chain of trapped ions that the search process can be strongly
improved when interferences are observed than employing a one by
one search. The speed up of the search process results from the
fact that the interference patterns are produced by the light of
all emitting ions of the chain at once so that there is a
simultaneous contribution of all scatterers to the signal. This
superposition of the signal is apparently sufficient for the speed
up and no entanglement between the ions nor other quantum
phenomena are required \cite{grover2}. However, quantum
interferences allow to increase the parameter space available and
in particular improve the precision when a larger number of ions
is involved, since they relax the demands for the spatial
resolving power of the detectors. Moreover, in the case of quantum
interferences an interference pattern is obtained even at ion
distances smaller than $\lambda/2$ which is the ultimate limit for
a classical interference experiment.
\par
{\bf Acknowledgement.} One of the authors, (G.O.A.) gratefully
acknowledges the support of the Alexander von Humboldt Foundation
Fellowship.

\end{document}